# Ultrasonic preparation of mesoporous silica using pyridinium ionic liquid

A. M. Putz[1], A. Len[2], C. Ianăşi[1], C. Savii[1], L. Almásy[2]*

[1] *Institute of Chemistry Timisoara of Romanian Academy, Laboratory of Inorganic Chemistry, Bv. MihaiViteazul, No.24, RO-300223 Timisoara, Romania*

[2]*Wigner Research Centre for Physics, Institute for Solid State Physics and Optics, POB 49 Budapest-1525, Hungary*

**Graphical abstract**

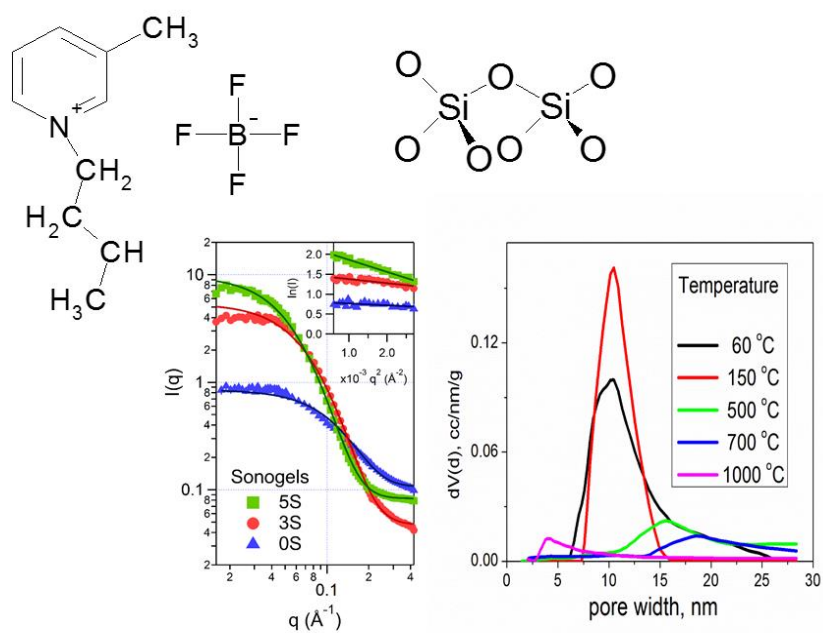

**Abstract**

Mesoporous silica matrices have been prepared via classic acid catalyzed and sono-catalyzed sol-gel routes. Tetramethoxysilan (TMOS) and methyl-trimethoxysilane (MTMS) were used as silica precursors, and N-butyl-3-methylpyridinium tetrafluoroborate ([bmPy][$BF_4$]) was employed as co-solvent and pore template. The ionic liquid (IL) to silica mole ratio was varied between 0.007 and 0.07. Nitrogen adsorption-desorption and small-angle neutron scattering measurements have been used to characterize the obtained materials. The ionic liquid played the role of catalyst that affected the formation of the primary xerogel particles, and changed the porosity of the materials. Ultrasound treatment resulted in microstructure change on the level of the colloid particle aggregates. In comparison with IL containing xerogels, the IL containing sonogels show increased pore diameter, bigger pore volumes and diminished surface areas.

**1. Introduction**

Ionic liquids (ILs) are salts that are liquid below 100 °C and consist of a large organic cation together with an organic or inorganic anion [1,2]. ILs can be used as reaction medium for inorganic materials due to their intrinsic high charge density and polarizability [3,4]. ILs pre-organised structure makes them suitable as templates for the porous materials [3,4]. ILs used as templating agent can increase the thermal stability of the synthesized materials [5]. As solvents in sol-gel process, ILs negligible vapour pressure prevents the evaporation and their high ionic strength increases the speed of aggregation [6]. Shrinkage and syneresis are reduced during gelation and aging, therefore due to a limited collapse of the pores during solvent extraction, highly porous silica matrixes can be synthesized [7,8]. Besides their role as co-solvents, ILs also show catalytic activity. The pyridinium groups of the cations behave as Brønsted acidic catalysts that accelerate the hydrolysis of silica precursors; while the tetrafluoroborate anion behaves as Lewis base catalysts that accelerate the condensation reactions of silica [9].

Use of methylated precursors leads to water stable and time stable ionogels [10,11] and improves immobilization of substituted porphyrins [12]. Besides inducing hydrophobicity, the bulky IL cations also help to control the pore size [5]. It was shown that [bmPy][BF$_4$] or [bmIm][BF$_4$] (at molar ratio IL /Si =1.6 and IL/Si =0.07, respectively) causes a significant increase in the average pore radius, in comparison with similar materials prepared without IL [4].

Sonication is an alternative method to stimulate the hydrolysis of the alkoxide without using alcoholic solvents in the sol-gel process [13,14]. Strong ultrasound action (20-100 kHz) induces chemical reactions and/or enhances their rate [15]. Soft sonication augments the aging of the starting solutions [16,17].

In the present study nanostructured xerogels and sono-activated gels have been prepared by using N-butyl-3-methylpyridinium tetrafluoroborate [bmPy][BF$_4$] ionic liquid (IL) as co-solvent, at IL/Si molar ratios from 0.007:1 to 0.07:1. Tetramethoxysilan (TMOS) and methyltrimethoxysilane (MTMS) were used as precursors; TMOS was chosen because it leads to faster hydrolysis in comparison with TEOS [18]. The obtained materials were characterised by FT-IR spectroscopy, nitrogen adsorption and small-angle neutron scattering.

## 2. Experimental
### 2.1. Synthesis conditions

All reagents were used as received: N-butyl-3-methylpyridinium tetrafluoroborate ([bmPy][BF$_4$], for synthesis 98%, Merck); the silicon precursors: methyl-trimethoxysilane (MTMS, 98%, Merck) and tetramethoxysilane (TMOS, 98%, Merck); distilled water; and hydrochloric acid (HCl, 37%, S.C. SilalTreding SRL). MTMS to TMOS molar proportion of 1:8 was used for all samples.

Two series of samples: classic xerogels and sonicated gels were prepared using a modified recipe of Karout and Pierre [4]. The synthesis started by mixing 0.13 moles TMOS, 1.15 moles distilled water, 0.4 moles acidic water (adjusted to pH= 2.8 by addition of HCl) and 0.01615 moles MTMS.

In the case of xerogel series, the reaction mixture was mechanically vigorously stirred for 15 minutes in a round bottom flask until a homogenous sol was obtained. At this point, for each sample, aqueous [bmPy][$BF_4$] solution was added into the reaction mixture in various IL/Si molar ratios, within 0.007 to 0.07 range, (see Table 1) under continuous stirring for 15 minutes. The obtained sols were left to gel, in covered vessels, at room temperature. The xerogel samples labelled as 0X-60, 1X-60, 2X-60, 3X-60, 4X-60 and 5X-60 were obtained by drying the silica gel at 60 °C for 26 h, see Table 1. For 0X-60, instead of the aqueous ionic liquid solution, an equal volume of distilled water was added.

A second series of samples, 0S-60, 1S-60, 2S-60, 3S-60, 4S-60 and 5S-60, was synthesized by using the same recipe, but instead of mechanical stirring, ultrasonication was applied starting from adding of the first reactant into the reaction mixture, till the end.

A VCX-T – 750 apparatus operating at 20 kHz and 750 W, with titanium sonotrode (1 cm diameter) was immersed (1cm deep) into the sol. The sonotrode worked at 25% amplitude, with pulse sequence on/off - 15/5 sec. A 150 mL sample container was filled with 95 mL of liquid for each preparation. During sonication (23 minutes) the temperature increased from 25 ºC to 60 ºC. The total transferred energy, including thermal dissipation, was 250 J/mL for each sample.

**2.2. Thermal treatment**

The 0X-60, 5X-60 and 5S-60 samples were thermally successively treated, in air, at 150 °C for 90 minutes (labelled 0X-150, 5X-150, 5S-150); at 500 °C for 60 minutes (labelled 0X-500, 5X-500, 5S-500); and at 1000 °C for 60 minutes, labelled as 0X-1000, 5X-1000 and 5S-1000.

Table 1. Synthesis parameters

| Xerogel samples | Sonogel samples | Molar ratio IL/Si | IL [g] | TMOS [g] | MTMS [g] |
|---|---|---|---|---|---|
| 0X | 0S | 0 | 0 | 19.8 | 2.2 |
| 1X | 1S | 0.007 | 0.25 | 19.8 | 2.2 |
| 2X | 2S | 0.018 | 0.62 | 19.8 | 2.2 |
| 3X | 3S | 0.034 | 1.19 | 19.8 | 2.2 |
| 4X | 4S | 0.053 | 1.85 | 19.8 | 2.2 |
| 5X | 5S | 0.07 | 2.46 | 19.8 | 2.2 |

**2.3. Materials Characterisation**

FT-IR spectra of the powders in KBr pellets were registered on a JASCO –FT/IR-4200 apparatus.

Textural properties of the composite silica matrix samples were analyzed using nitrogen adsorption/desorption measurements at liquid nitrogen temperature (77K) with a Quantachrome Nova 1200e instrument. Prior to the measurements the samples were degassed during 3 hours at 100ºC, in flow nitrogen atmosphere. The specific surface area ($S_{BET}$) was calculated from the Brunauer, Emmett and Teller (BET) equation, the total pore volume ($V_p$) determined from the last point of adsorption and the pore diameter ($D_p$) determined by Barrett, Joyner and Halenda (BJH) method, from the adsorption and desorption branches of the isotherms, using a NovaWin software. The pore size distribution was calculated by the DFT method.

The small angle neutron scattering (SANS) measurements were performed at the *Yellow Submarine* spectrometer at the Budapest Neutron Centre [19]. Samples have been measured at ambient temperature.

## 3. Results and Discussion

### 3.1. Gelation time

As IL content was increasing, the gelation time was decreasing, specifically, from 24h to 5h for the xerogel series and from 126h to instant gelation for the sonogel series; therefore the catalytic role of IL was evidenced.

### 3.2. FT-IR Analysis

The broad band between 3474–3441 cm$^{-1}$ corresponds to the overlapping of the O-H stretching bands of hydrogen-bonded water molecules (H-O-H. . .H) and SiO-H stretching of surface silanols hydrogen-bonded to molecular water (SiO-H…H$_2$O) [20,21]. This band is more intense in the sample synthesised without IL, in both xerogels (Figure 1) and the calcined samples (Figure 2), indicating that IL's presence strengthens the hydrophobic character. The adsorbed water deformation vibration bands [21,22] appear at 1635 cm$^{-1}$ as well.

The intense and broad band which appears at 1099–1075 cm$^{-1}$ is assigned to the transversal optical (TO) modes of the Si-O-Si asymmetric stretching vibrations [21,23]. The presence of the silanol groups were confirmed by the band centred about 960 cm$^{-1}$, which is associated with the stretching mode of the Si–OH groups [24]. The symmetric stretching vibrations of Si-O-Si appear at 800 cm$^{-1}$ and 680 cm$^{-1}$ [20,21,25] and its bending mode appears at 467–458 cm$^{-1}$ [21,24,26,27].

The band from 2970-2975 cm$^{-1}$ corresponds to the asymmetric C-H stretching vibration of methyl group from the IL [28] and its disappearance after calcinations at 500°C shows that the bmPy cation is largely removed.

Adsorption bands of the IL were not observed in our samples. The characteristic 1027 cm$^{-1}$ band of pure [bmPy][BF$_4$] overlapped with the broad band of the silica matrix at 1084 cm$^{-1}$ and could not be distinguished [29]; The Si-O-Si asymmetric stretching vibrations were shifted to slightly higher wave numbers with the increase of calcinations temperature, while the symmetric stretching vibrations and their bending mode were shifted to slightly lower wave numbers. With the increase of the calcinations temperature (from 150 to 500°C), the bands assigned to antisymmetric deformation bending of $CH_2$ groups [shifted from 1509 $cm^{-1}$ (60C and 150C IL samples), to 1396 $cm^{-1}$ (for 500C IL samples]. These changes indicate a restructurization of the silica network following the elimination of the ionic liquid and the pore collapse (shown by nitrogen adsorption). The 1640 and 960 bands characteristic for the Si-OH become weaker, indicating water loss. The IR data did not show visible differences between xerogels and sonogels (see Figures 1 and 2).

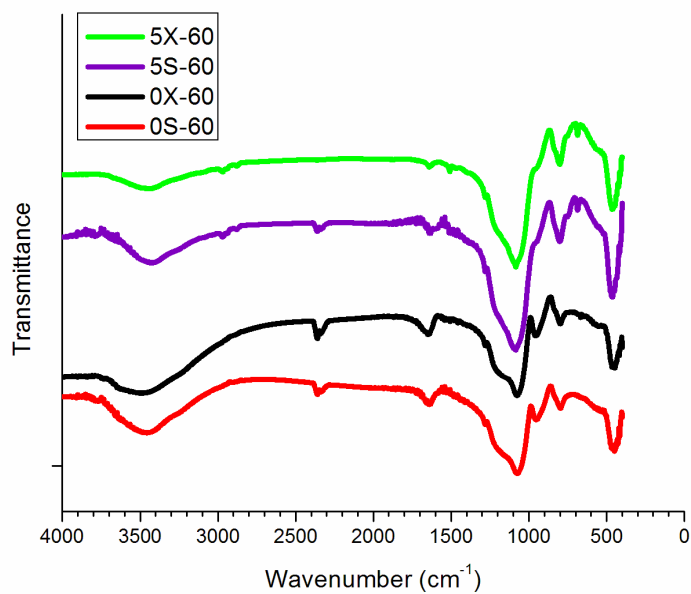

*Figure 1. IR spectra of the xerogels 0X-60, 1X-60, 5X-60, and sonogels 0S-60, 1S-60, 5S-60.*

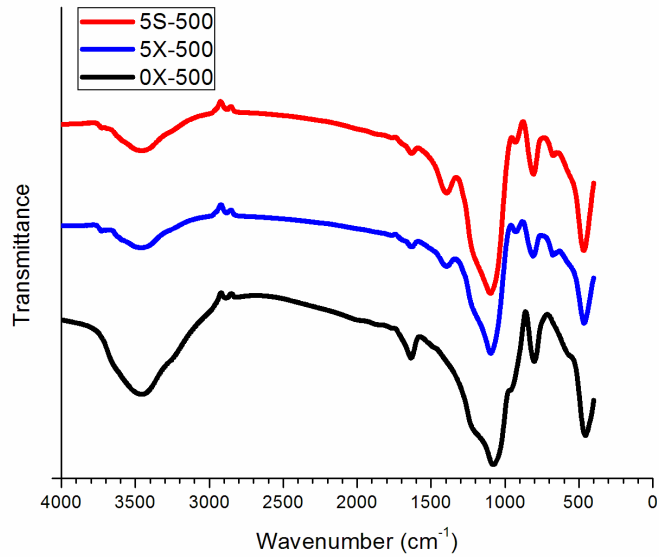

*Figure 2. IR spectra of the calcined samples 0X-500, 5X-500 and 5S-500.*

**3.3 Nitrogen adsorption-desorption isotherms**

It is well known that increased surface tension causes pore collapsing, therefore the reduction of the surface tension by the IL content is favourable for porosity preservation [4]. As a general observation, within both the xerogels and the sonogels series, the pore size was continuously increasing with increasing IL concentration (Table 2). At small IL content (1X-60 and 2X-60 samples) the values of the pore diameters from adsorption are close to those calculated from desorption, which indicates more regular pore shapes in these samples. In xerogel and sonogel synthesized without IL regular pore shapes are obtained as well.

The mean pore size, calculated from the desorption branch of isotherms using BJH method, increases from 3.64 nm (0X-60 sample) to 4.68 nm (5X-60 sample) and from 3.61 nm (0S-60 sample) to 7.87 nm (5S-60 sample) for the sonogels, see Table 3. Similar trends are obtained from the adsorption branches and by DFT method (see Table 2). The pore size in sonogel (5S-60 sample) is almost double compared to the corresponding xerogel (5X-60 sample).

The specific surface area is decreasing with the increase of IL concentration, in both, xerogels and sonogels, after a sharp step-like decrease between the 0X-60 and 1X-60, (and 0X-60 and 1S-60) samples.

With the increase of calcinations temperature, the specific surface area continuously decreases in both series until 700 °C. The pore diameter increases in both, xerogels and sonogels, as IL was removed. The surface area reduction during calcination was stronger for the xerogel series compared to sonogel series, indicating that sonogels were more resistant to calcinations.

Table 2. Textural properties of the xerogels and sonogels

| Sample | Surface Area BET [m²/g] | Pore Diameter [nm] Ads | Pore Diameter [nm] Des | Total pore volume [cm³/g] |
|---|---|---|---|---|
| **0X-60** | 774.8 | 3.62 | 3.64 | 0.424 |
| **1X-60** | 535.7 | 3.62 | 3.63 | 0.452 |
| **2X-60** | 446.9 | 3.61 | 3.67 | 0.552 |
| **3X-60** | 389.8 | 4.62 | 3.65 | 0.542 |
| **4X-60** | 308.2 | 6.18 | 4.56 | 0.509 |
| **5X-60** | 248.0 | 4.62 | 4.682 | 0.397 |
| **0S-60** | 943.7 | 3.61 | 3.61 | 0.709 |
| **1S-60** | 542.6 | 3.39 | 3.86 | 0.523 |
| **2S-60** | 373.0 | 5.58 | 4.34 | 0.554 |
| **3S-60** | 390.2 | 5.58 | 4.93 | 0.592 |
| **4S-60** | 172.1 | 7.67 | 6.61 | 0.362 |
| **5S-60** | 214.0 | 9.038 | 7.87 | 0.682 |

Table 3. Textural properties of the thermally treated samples

| Sample | Surface Area BET [m²/g] | Pore Diameter [nm] Ads. | Pore Diameter [nm] Des. | Pore Diameter [nm] DFT | Total pore volume [cm$^3$/g] |
|---|---|---|---|---|---|
| **5X-60** | 248.0 | 4.238 | 4.682 | 5.880 | 0.397 |
| **5x-150** | 215.7 | 6.218 | 4.682 | 6.079 | 0.371 |
| **5x-500** | 94.91 | 15.191 | 8.955 | 18.55 | 0.448 |
| **5x-700** | 71.93 | 15.716 | 17.82 | 17.91 | 0.329 |
| **5x-1000** | 57.34 | 3.717 | 17.57 | 3.418 | 0.165 |
| **5s-60** | 214.0 | 15.287 | 9.038 | 9.773 | 0.682 |
| **5s-150** | 188.5 | 15.150 | 8.959 | 9.773 | 0.647 |
| **5s-500** | 103.3 | 4.278 | 16.27 | 3.418 | 0.425 |
| **5s-700** | 61.51 | 3.726 | 25.24 | 27.37 | 0.218 |
| **5s-1000** | 63.34 | 3.656 | 3.58 | 3.934 | 0.118 |

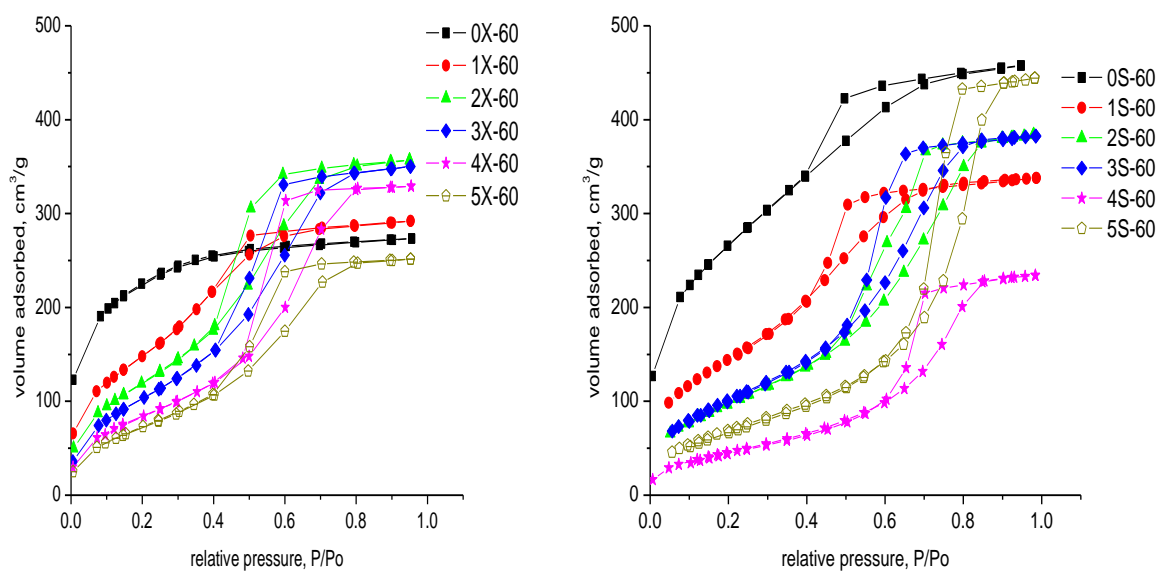

Figure 3. *Nitrogen adsorption/desorption isotherms of the xerogel and sonogel samples*

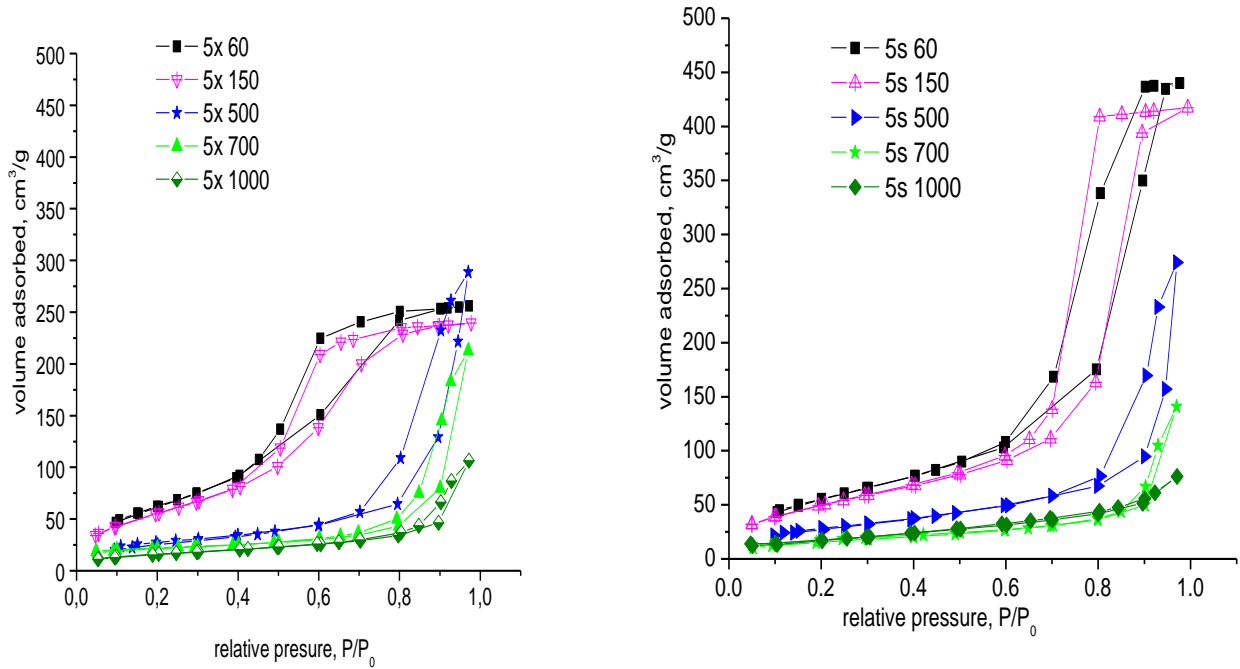

Figure 4. *Nitrogen adsorption/desorption isotherms of the xerogel and sonogel samples calcined at different temperatures*

### *3.3.1. Nitrogen adsorption-desorption isotherms for xerogels and sonogels dried at 60ºC*

Nitrogen adsorption isotherms of samples dried at 60ºC are shown in Figure 3. The xerogel sample without IL, 0X-60, has Type IV isotherms with H4 hysteresis loop which is often associated with narrow slit-like pore [30]. The xerogel samples synthesized with IL, have Type IV isotherm with H2 characteristic hysteresis loops attributable to the materials with ink-bottle pore shapes [31]. The pore diameters from adsorption reflect the size of the bottle, and the pore diameters from desorption can be associated with the necks between pores [32]. The sonogel sample synthesized without IL, 0S-60, and the sonogels samples with IL content up to 0.053 mole fraction present Type IV isotherm with H2 characteristic hysteresis loops. The sonogel sample with the highest concentration of IL, 5S-60, has Type IV isotherm with H1 characteristic hysteresis loops, associated with porous materials which exhibit a narrow distribution of relatively uniform pores[31].

*3.3.2. Nitrogen adsorption-desorption isotherms for the thermally treated samples*

Isotherms of the thermally treated samples are shown in Figure 4. For the xerogel series, sample 5X-150 has Type IV isotherm with H2 hysteresis loop. By heating at 500 °C, 700 °C and 1000 °C, it changes to Type IV isotherms with H3 hysteresis loops, characteristic for slit shaped pores [33]. These data suggest that in order to obtain uniform pores the calcination temperatures should be below 500°C.

In the case of sonogels, Type IV isotherm with H1 hysteresis loops were obtained both for the sample dried at 60 °C, as well as after thermal treatment at 150 °C. Further calcination at 500 °C and 700 °C leads to Type IV isotherms with H3 hysteresis loops, similar to the xerogel samples, and at 1000 °C the Type II isotherms indicate pores collapse [33]. The thermally treated sonogels, in comparison with the corresponding xerogel samples, showed larger pore sizes, larger total pore volumes and reduced surface areas. This is in good agreement with a recent report dealing with ultrasonication effects in xerogels with high IL content [34].

*3.3.3. Pore size distribution*

Narrow size distributions were obtained only for the xerogel and sonogel samples, dried at 60 °C and thermally treated at 150 °C, see Figure 5. The PSD is slightly broader and the pores are larger in the sonogels. With the increase of the calcinations temperature, above 500 °C, the PSD becomes broader, as a result of disruption of the walls between some pores, leading to overall pore size increase. Calcination temperatures over 500 °C are too high for obtaining materials with narrow size distribution. PSD reveal also that the pore sizes of the present materials are in the mesoporous range. Verma and co-workers also obtained a more uniform pore size distribution for sonogels in comparison with the xerogels [34].

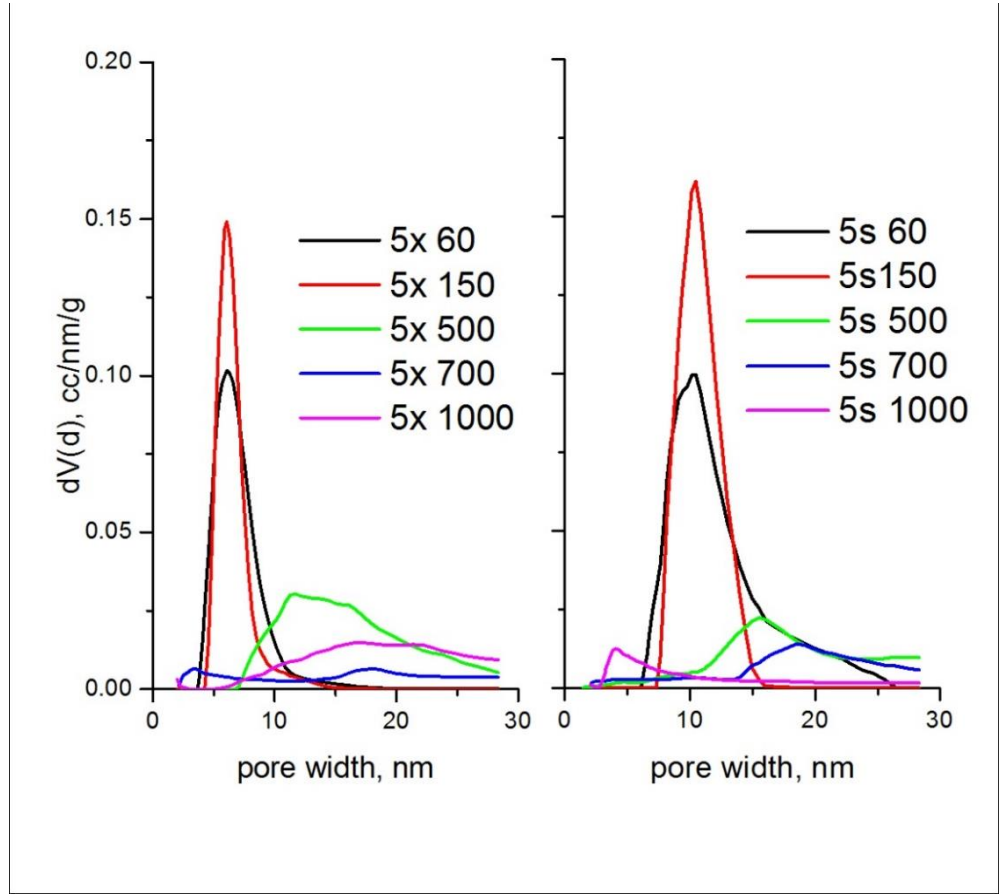

*Figure 5. Pore size distributions calculated by DFT of the thermally treated samples.*

## 3.4 SANS

The scattering intensity curves for xerogels and sonogels are shown in Figure 6. They are characteristic for a material with inhomogeneities having a characteristic size, as seen by the broad plateau with decay starting at higher $q$ values. These can be attributed to the primary particles that form in the initial phase of the condensation reaction. The simplest estimation of the particle size is the Guinier equation which could be used on the small $q$ part of the data (inset in Figure 6). Compact particles with rough, fractal-like surface can be modelled by Eq. (1):

$$I(q) = \frac{K}{(1+q^2\xi^2)^p} + bg \qquad (1),$$

where *K* is a scale factor related to the number density and contrast of the particles, *ξ* is a characteristic length related to the particle size and *p* is related to the surface roughness. The *bg* term contains the incoherent background coming mostly from the hydrogen atom content of the materials. The radius of gyration *Rg* and the correlation length *ξ* are not necessarily proportional to each other because the exponent *p* changes from case to case, but both can be considered as an approximate measure of the size of the scattering entity [35]. Interparticle interference effects are not included in this model, as in the given *q*-range there is no visible decrease or increase of the scattering intensity toward low *q* values. It is known that at a larger magnification, (that is at lower *q*) the fractal like assembly of the primary particles would be seen [36,37]. Data fitting was performed by the NCNR reduction and analysis package written for the Wavemetrics Igor Pro software [38].

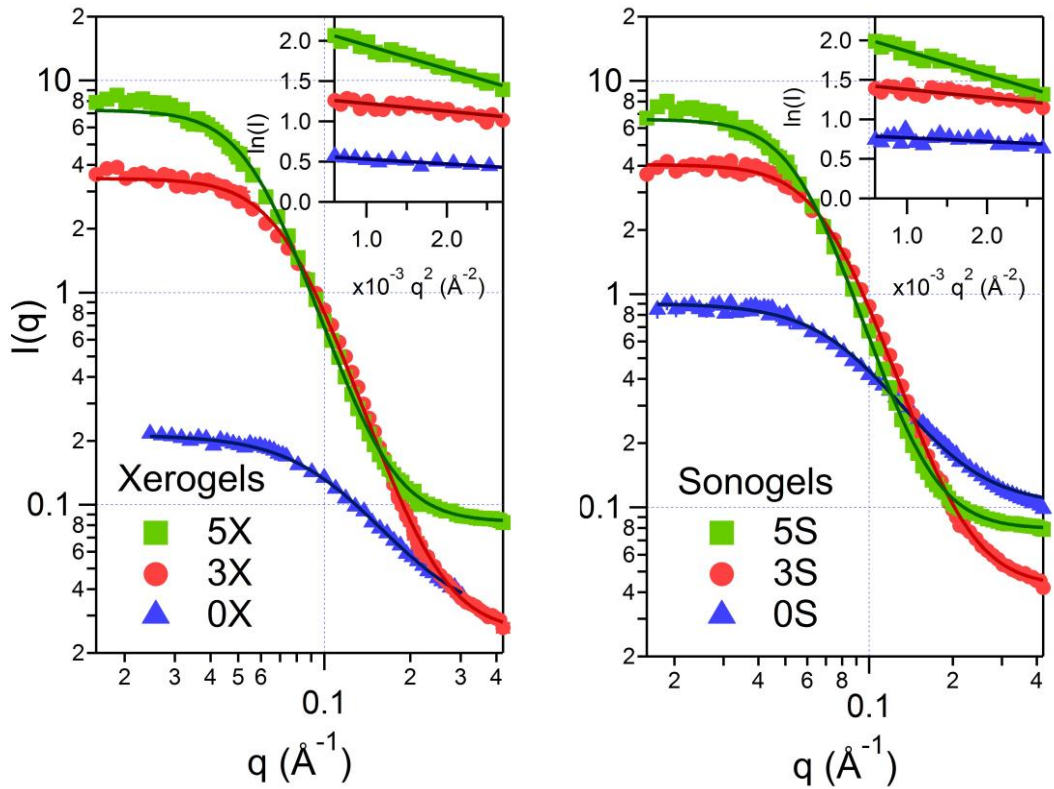

*Figure. 6.*

*SANS scattering curves of xerogels (left) and sonogels (right). Continuous lines represent fitting by Eq. 1. In the Inset the Guinier fits are shown; here the data are shifted vertically for better view.*

The scattering data show that the primary particle size in the xerogels and sonogels is about 2 nm, while at higher ionic liquid content they grow to 4-6 nm (Figure 7). No appreciable difference could be seen between the structure of sonogels and xerogels.

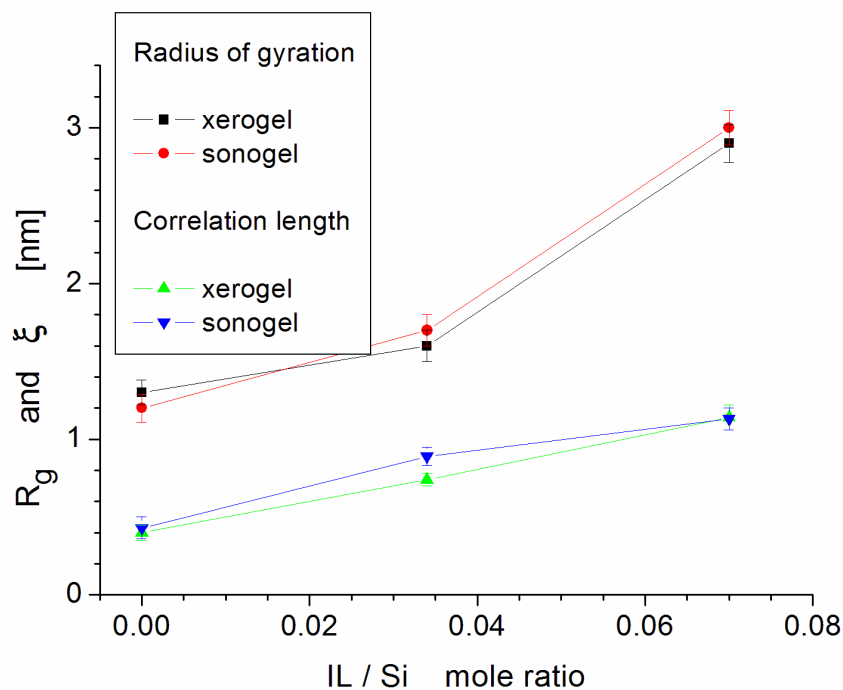

*Figure 7. Radii of gyration and correlation lengths, obtained from the SANS data.*

In general, sonication applied during synthesis does not affect the size of monomer particles but may influence their secondary aggregation [36,37,39]. In the present materials, the SANS data did not reveal any influence of ultrasonication on the level of the primary particle formation, while the presence of ionic liquid appears to slow down the polycondensation reaction thus leading to increase of the primary particle size.

**4. Conclusions**

Xerogels containing short chain ionic liquid, have been prepared by ultrasonication and characterized by a variety of techniques. The ionic liquid used as co-solvent played both catalytic and templating role, as shown by the decrease of gelation time and influence on the development of different porosity. Application of ultrasound influenced the pore size distribution, surface area and pore volume. Nitrogen adsorption data show that the ultrasonic field action during the gelification leads to larger pore sizes, bigger pore volumes and smaller surface areas in the calcined materials. The pore size is almost double for sonogel samples with the highest studied IL content, compared to the non-sonicated xerogels. Increasing calcination temperatures caused decrease of the specific surface areas due to the sintering which followed the IL elimination. The primary particle sizes do not differ significantly upon applying ultrasound, but the addition of IL slows down the polycondensation and produces particles of larger size.

ACKNOWLEDGEMENTS

Authors thanks Romanian Academy. Neutron scattering experiments from this research project has been supported by the European Commission under the 7$^{th}$ Framework Programme through the key action: Strengthening the European Research Area, Research Infrastructures. Grant Agreement No 283883-NMI3-II.